\documentclass[fleqn,usenatbib]{mnras}

\usepackage{newtxtext,newtxmath}

\usepackage[T1]{fontenc}

\DeclareRobustCommand{\VAN}[3]{#2}
\let\VANthebibliography\thebibliography
\def\thebibliography{\DeclareRobustCommand{\VAN}[3]{##3}\VANthebibliography}

\usepackage{graphicx}	
\usepackage{amsmath}	

\newcommand{\Ni}{\ensuremath{^{56}\mathrm{Ni}}}
\newcommand{\Co}{\ensuremath{^{56}\mathrm{Co}}}

\newcommand{\Msun}{\ensuremath{\mathrm{M}_\odot}}
\newcommand{\Msunpyr}{\ensuremath{\Msun~\mathrm{yr^{-1}}}}
\newcommand{\kmps}{\ensuremath{\mathrm{km~s^{-1}}}}

\graphicspath{{./}{figures/}}

\title[SN~Ia and wind interaction]{
Early excess emission in Type~Ia supernovae from the interaction between supernova ejecta and their circumstellar wind
}

\author[T. J. Moriya et al.]{
Takashi J. Moriya,$^{1,2}$\thanks{E-mail: takashi.moriya@nao.ac.jp (TJM)}
Paolo A. Mazzali,$^{3,4}$
Chris Ashall,$^5$
and
Elena Pian$^6$
\\
$^{1}$National Astronomical Observatory of Japan, National Institutes of Natural Sciences, 2-21-1 Osawa, Mitaka, Tokyo 181-8588, Japan \\
$^{2}$School of Physics and Astronomy, Faculty of Science, Monash University, Clayton, Victoria 3800, Australia \\
$^{3}$Astrophysics Research Institute, Liverpool John Moores University, IC2, Liverpool Science Park, 146 Brownlow Hill, Liverpool L3 5RF, UK\\
$^{4}$Max-Planck Institute for Astrophysics, Karl-Schwarzschild-Stra{\ss}e 1, 85748 Garching, Germany \\
$^{5}$Department of Physics, Virginia Tech, 850 West Campus Drive, Blacksburg VA, 24061, USA \\
$^{6}$INAF, Astrophysics and Space Science Observatory, via P. Gobetti 101, 40129 Bologna, Italy \\
}

\date{Accepted 2023 May 04. Received 2023 April 15; in original form 2023 February 08}

\pubyear{2023}

\begin{document}
\label{firstpage}
\pagerange{\pageref{firstpage}--\pageref{lastpage}}
\maketitle

\begin{abstract}
The effects of the interaction between Type~Ia supernova ejecta and their
circumstellar wind on the photometric properties of Type~Ia supernovae are
investigated. We assume that a hydrogen-rich, dense, and extended circumstellar
matter (CSM) is formed by the steady mass loss of their progenitor systems. The
CSM density is assumed to be proportional to $r^{-2}$. When the
mass-loss rate is above $10^{-4}~\Msunpyr$ with a wind velocity of 100~\kmps,
CSM interaction results in an early flux excess in optical light-curves within
4~days of explosion. In these cases, the optical colour quickly
evolves to the blue. The ultraviolet flux below 3000~\AA\ is found to have a
persistent flux excess compared to Type~Ia supernovae as long as CSM interaction
continues. Type~Ia supernovae with progenitor mass-loss rates between $10^{-4}$ and
$10^{-3}~\Msunpyr$ may not have a CSM that is dense enough to affect spectra to
make them Type~Ia-CSM, but they may still result in Type~Ia supernovae with an
early optical flux excess. Because they have a persistent ultraviolet flux
excess, ultraviolet light curves around the luminosity peak would be
significantly different from those with a low-density CSM.
\end{abstract}




\section{Introduction}\label{sec:introduction}

Type~Ia supernovae (SNe~Ia) are thermonuclear explosions of white dwarfs
\citep[e.g.,][]{nugent2011,bloom2012}. The evolutionary path leading to white
dwarf explosions is a long-standing issue in astrophysics \citep[e.g.,][for
recent reviews]{maeda2016,livio2018,wang2018,soker2019}. Two major paths have
been proposed. One is a so-called ``single-degenerate'' channel in which a white
dwarf gains its mass through the accretion from a non-degenerate companion star
\citep[e.g.,][]{nomoto1982}. The other path is a ``double-degenerate'' channel
in which coalescence of binary white dwarfs leads to a thermonuclear explosion
\citep[e.g.,][]{iben1984,webbink1984}. 

The two channels predict different properties of circumstellar matter (CSM)
around SNe~Ia. In the case of the single-degenerate channel, mass transfer from
a non-degenerate star to a white dwarf can lead to mass loss \citep[e.g.,][and
references therein]{chomiuk2012,moriya2019}. Thus, single-degenerate systems
can lead to the formation of an extended dense hydrogen-rich or helium CSM
\citep[e.g.,][]{dragulin2016}. However, the double-degenerate channel involves
two white dwarfs. Therefore, no significant mass loss forming a dense extended
CSM is expected from such a system before the explosion. Observationally, a
fraction of SNe~Ia show clear signatures of interaction between the SN~Ia ejecta
and an extended dense hydrogen-rich or helium CSM in their optical light-curves
and spectra
\citep[e.g.,][]{hamuy2003,dilday2012,silverman2013,kool2022,sharma2023}, and
they are likely related to the single-degenerate channel. It is also possible
that SN~Ia signatures are often hidden in the interaction features and a certain
fraction of interacting SNe are linked to SNe~Ia
\citep[e.g.,][]{leloudas2015,jerkstrand2020}. The fact that some SNe~Ia tend to
appear in star-forming regions also suggests importance of large mass loss in
their progenitor systems and implies the existence of dense CSM around SNe~Ia
\citep[e.g.,][]{bartunov1994,anderson2015,pavlyuk2016,hakobyan2016}.

Several observational features have been linked to the possible presence 
of hydrogen-rich dense CSM around SNe~Ia. For example, strong calcium absorption
observed in early spectra of SNe~Ia has been related to the existence of
hydrogen-rich CSM
\citep[e.g.,][]{gerardy2004,mazzali2005a,mazzali2005b,tanaka2008}. Some SNe~Ia
also show hydrogen emission lines at late phases that could be excited by CSM
interaction
\citep[e.g.,][]{kollmeier2019,vallely2019,prieto2020,elias-rosa2021}. In
addition, narrow, blueshifted sodium absorption lines detected in SNe~Ia are
suggested to indicate the existence of extended CSM around their progenitors
\citep[e.g.,][]{patat2007,maguire2013}. The density of extended CSM around SN~Ia
progenitor systems can also be probed by radio and X-ray observations of SNe~Ia.
However, no radio and X-ray emission from SNe~Ia has been observed
\citep[e.g.,][]{chomiuk2012,chomiuk2016,perez-torres2014,harris2016,harris2018,harris2021,lundqvist2020,russell2012,margutti2012,margutti2014,dwarkadas2022}
except for SN~2020eyj \citep{kool2022}\footnote{There is a controversial case of
SN~2005ke \citep{immler2006,hughes2007}. SN~2012ca was detected in X-rays \citep{bochenek2018}, but it may not be a SN Ia \citep{inserra2016}.}. The lack of radio and X-ray detection
indicates that the CSM density around SNe~Ia is generally not as high as those
predicted by the single-degenerate channel, but it is important to note that the
inferred CSM density depends on uncertain physics determining the emission
properties \citep[e.g.,][]{chomiuk2016}.
It is also possible that there is a delay between the accretion phase and explosion in the single-degenerate channel \citep[e.g.,][]{distefano2011,justham2011}.

In addition to the aforementioned CSM signatures, early emission from SNe~Ia is
suggested to contain information on progenitor systems of SNe~Ia. Recent early
SN~Ia observations are starting to reveal that some SNe~Ia have excess optical
emission within several days after the explosions
\citep[e.g.,][]{goobar2014,cao2015,marion2016,jiang2017,jiang2018,jiang2020,jiang2021,hosseinzadeh2017,hosseinzadeh2022,miller2018,miller2020,shappee2019,li2019,dimitriadis2019,tucker2021,burke2021,ashall2022,sai2022,ni2022,srivastav2022,lim2023}.
The fraction of SNe~Ia showing early excess emission is estimated to be
$\sim 20-30$\% \citep[e.g.,][]{deckers2022,magee2022}. Early excess emission was
originally suggested to be caused by the impact of the SN ejecta on a
non-degenerate companion star in the progenitor system, and therefore to
constitute evidence in favour of the single-degenerate channel
\citep[e.g.,][]{kasen2010,kutsuna2015,maeda2014,liu2015}. Other
mechanisms have been proposed to cause early excess emission, such as mixing of
\Ni\ to outer layers
\citep[e.g.,][]{piro2016,noebauer2017,maeda2018,polin2019,magee2020,magee2021},
Doppler line shifts \citep[][]{ashall2022}, and existence of carbon and oxygen-rich envelope around white dwarfs \citep[e.g.,][]{ashall2021,maeda2023}.
In addition to the excess
emission, early colour evolution is found to have diversity in SNe~Ia
\citep[e.g.,][]{stritzinger2018,bulla2020,burke2022}.

Another possible way to produce early excess emission is the interaction between
the SN ejecta and a locally produced CSM. Interaction between SN ejecta and a
confined, dense CSM composed mostly of carbon and oxygen has been suggested to
cause early excess emission in SNe~Ia
\citep[e.g.,][]{dessart2014,levanon2015,levanon2017,levanon2019,piro2016,kromer2016}.
This type of CSM is expected to form during the
merger of two white dwarfs in the double-degenerate channel
\citep[e.g.,][]{hachinger2012,schwab2012,shen2012,pakmor2013,tanikawa2015}. 
However, in the
case of the single-degenerate channel, extended hydrogen-rich CSM formed by the
mass loss from progenitor systems is expected to exist as discussed earlier. In
this study, we investigate how the early emission properties of SNe~Ia are
affected by the interaction between SN ejecta and such an extended hydrogen-rich
CSM.

The rest of the paper is organized as follows. We first show our numerical modelling setups and the initial conditions of our numerical simulations in Section~\ref{sec:methods}. We summarize our results in Section~\ref{sec:results}. We discuss our results in Section~\ref{sec:discussion} and conclude this paper in Section~\ref{sec:conclusions}.

\section{Methods}\label{sec:methods}

\subsection{Numerical modelling}\label{sec:numerical_modelling}

The interaction between SN ejecta and CSM involves the conversion of kinetic
energy of SN ejecta to radiation. Thus, it is necessary to treat hydrodynamics
and radiation transfer, as well as their mutual effects, simultaneously.  We use
the one-dimensional multi-frequency radiation hydrodynamics code \texttt{STELLA}
\citep{blinnikov1998,blinnikov2000,blinnikov2006} to investigate the
consequences of the interaction between SN ejecta and CSM in SNe~Ia.
\texttt{STELLA} has been used for light-curve modelling of SNe~Ia
\citep[e.g.,][]{blinnikov2006,woosley2007,kamiya2012,noebauer2017}, including
the interaction between SN~Ia ejecta and a confined, dense carbon-oxygen CSM
\citep[][]{noebauer2016}. In short, \texttt{STELLA} numerically treats the
hydrodynamic equations coupled with radiation, and implicitly solves the
time-dependent equations of the angular moments of the radiation intensity
averaged over a frequency bin by using the variable Eddington method.
\texttt{STELLA} calculates spectral energy distributions (SEDs) at each time
step. Multicolour light-curves can be obtained by convolving filter functions
with the SEDs. The standard 100 frequency bins in the SED calculations, i.e.,
from 1~\AA\ to $5\times10^4$~\AA\ on a log scale, are adopted in this study. 
\texttt{STELLA} assumes spherical symmetry in calculating hydrodynamic
and radiation properties. Although SNe~Ia and their CSM may not be spherical
\citep[e.g.,][]{liu2017}, our assumption of spherical symmetry would be good
enough to obtain the overall properties of CSM interaction features in SNe~Ia
\citep[cf.][]{cikota2019}. 

\begin{figure}
	\includegraphics[width=\columnwidth]{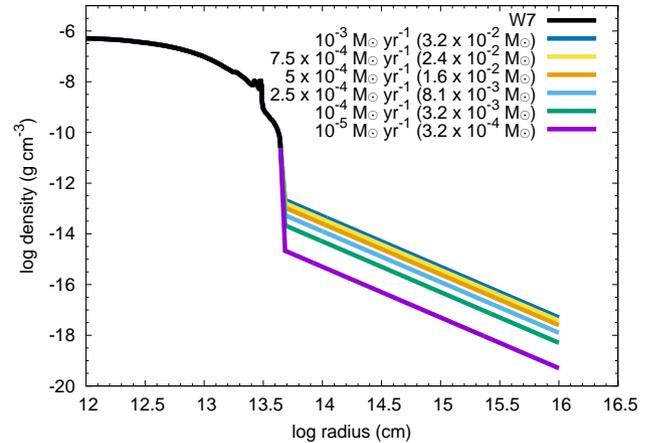}
    \caption{
    Initial density structure in our light-curve calculations. The CSM with $\rho_\mathrm{CSM}\propto r^{-2}$ are attached on top of the W7 density structure at $4\times 10^{13}~\mathrm{cm}$. The mass-loss rate of each model, as well as the mass in each CSM, is shown in the figure. We assume a wind velocity of 100~\kmps. 
    }
    \label{fig:initial_density}
\end{figure}

\begin{figure}
	\includegraphics[width=\columnwidth]{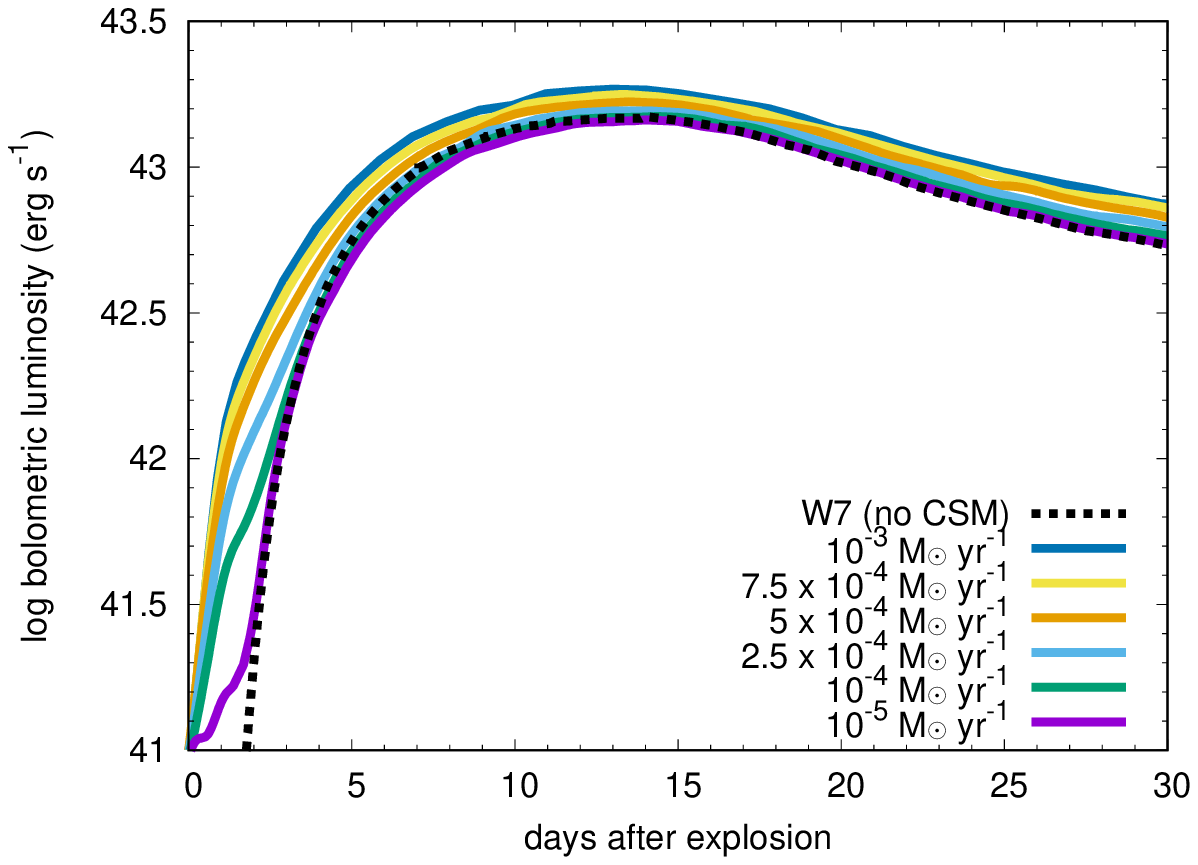}
	\includegraphics[width=\columnwidth]{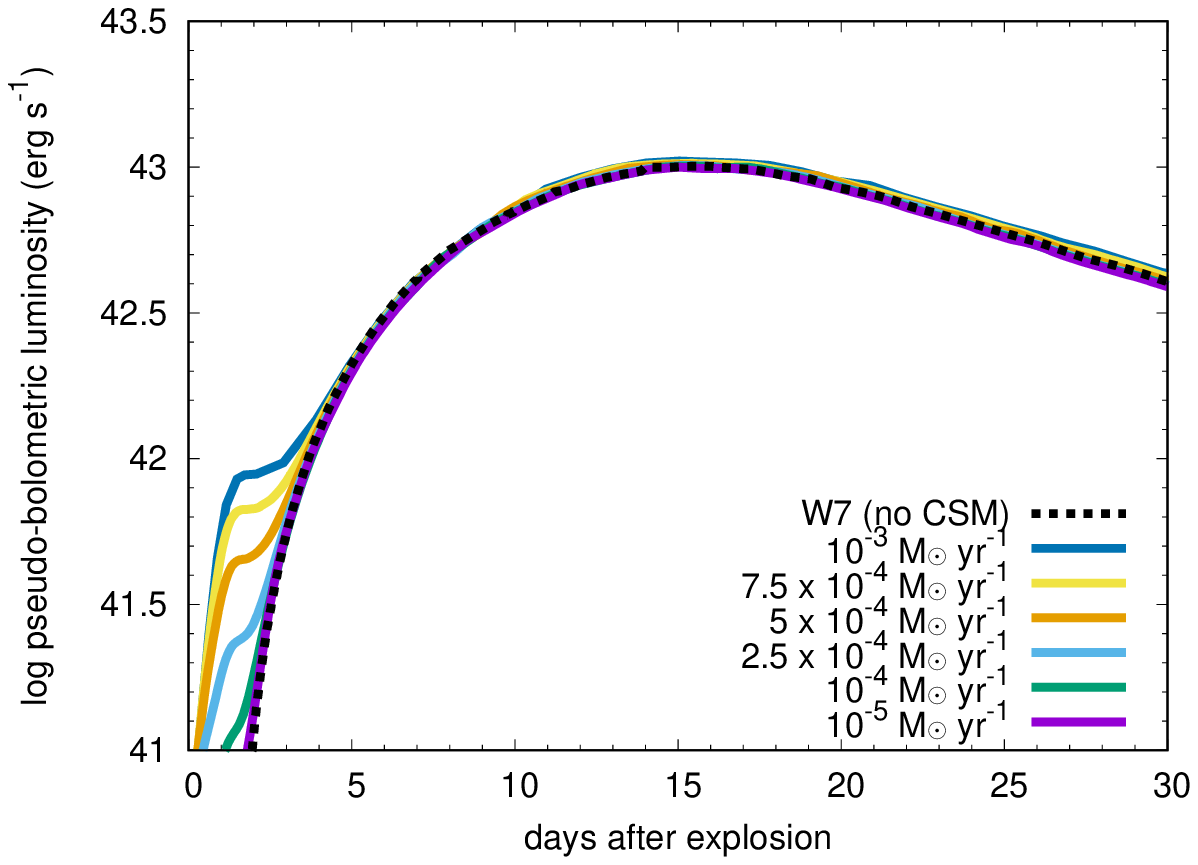}	
    \caption{
    Bolometric (top) and pseudo-bolometric (bottom) light-curves obtained from our numerical calculations. The bolometric light-curves include the flux in all the wavelengths and the pseudo-bolometric light-curves are obtained by integrating the optical flux from 3250~\AA\ to 8900~\AA.
    }
    \label{fig:bolometric}
\end{figure}

\begin{figure}
	\includegraphics[width=\columnwidth]{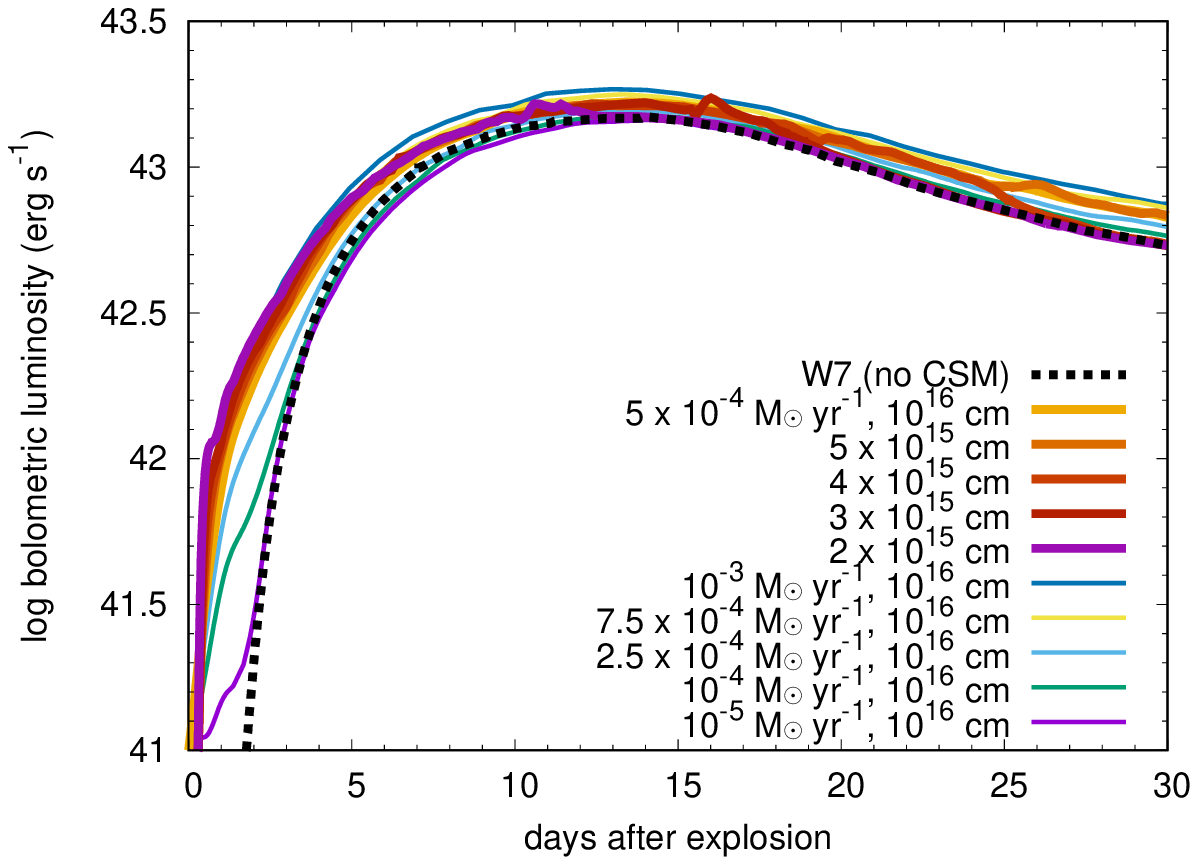}
	\includegraphics[width=\columnwidth]{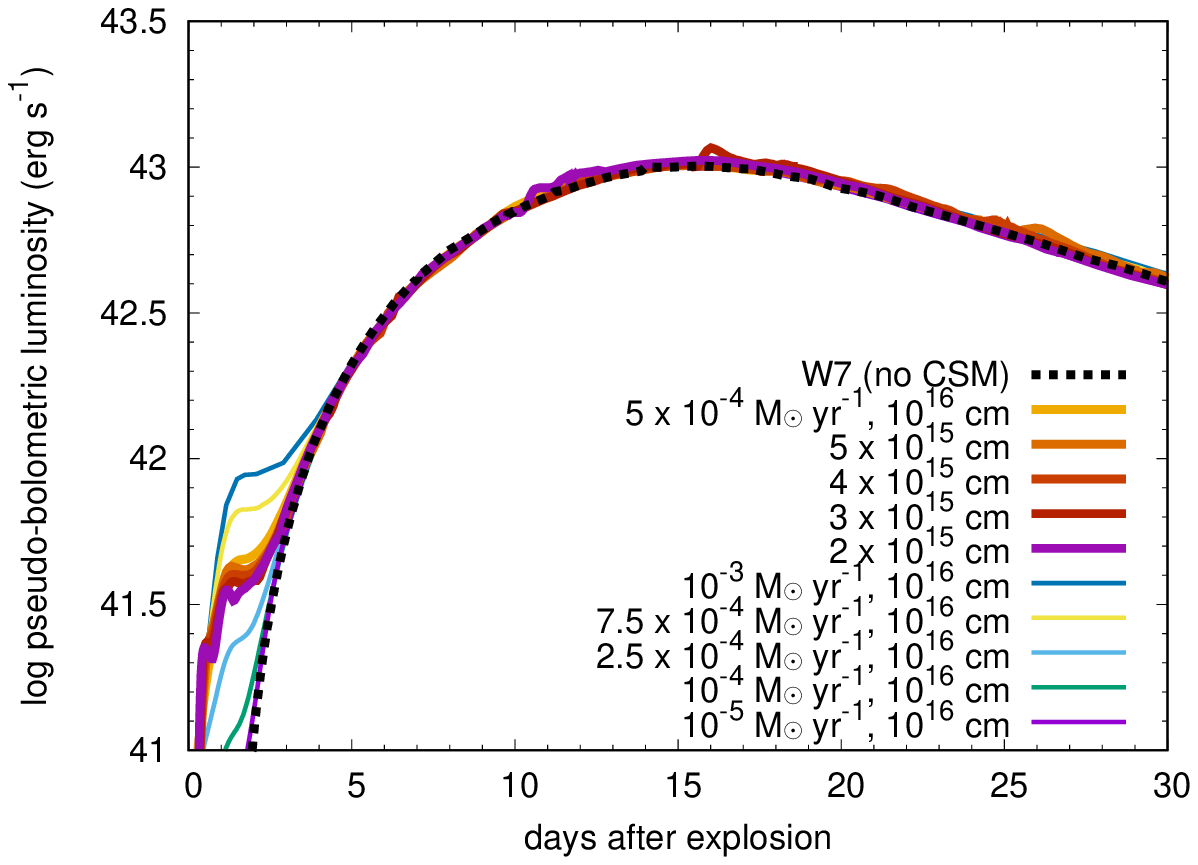}	
    \caption{
    Same as Fig.~\ref{fig:bolometric}, but we additionally show the models with different CSM radii in the case of $\dot{M}=5\times 10^{-4}~\Msunpyr$.
    }
    \label{fig:bolometric_radius}
\end{figure}

\subsection{Initial conditions}
\label{sec:initial_conditions}

We take the hydrodynamic structure of the W7 model in
\citet[][]{nomoto1984,thielemann1986} to represent the SN~Ia ejecta. The W7
model is a carbon fast deflagration explosion model of a white dwarf. The W7
model is known to explain the overall properties of SNe~Ia
\citep[e.g.,][]{nugent1997,tanaka2011}, although its nucleosynthesis is known to
have some issues \citep[e.g.,][]{iwamoto1999}. Because in the W7 model the SN
ejecta properties and the \Ni\ mass, which are the most influential photometric
properties of SNe, are overall consistent with SN~Ia observations \citep[e.g.,][]{stehle2005,mazzali2008}, it is
reasonable to adopt it to investigate the effects of CSM interaction in SN~Ia
photometric properties.

Before colliding on the CSM, the SN~Ia ejecta are assumed to expand homologously
for $2\times 10^4~\mathrm{sec}\simeq 0.2~\mathrm{days}$ when the outermost
layers reach $4\times 10^{13}~\mathrm{cm}$. Because the most extended
companion stars in SN~Ia progenitor systems are likely red giants having radii
of $\sim 10^{13}~\mathrm{cm}$ \citep[e.g.,][]{hachisu1996}, the CSM interaction
may actually start earlier. Still, the collision time is small compared to the
timescales of significant changes in photospheric properties. These occur on a
timescale of several days, as shown below, such that the assumed collision
radius is reasonable.

The CSM structure is attached to the SN ejecta at $4\times 10^{13}~\mathrm{cm}$
and extends above. If we assume that the progenitor system experienced steady mass
loss, the CSM density can be expressed as
\begin{equation}
    \rho_\mathrm{CSM} = \frac{\dot{M}}{4\pi v_\mathrm{wind}}r^{-2}, \label{eq:csmdensity}
\end{equation}
where $\dot{M}$ is the mass-loss rate, $v_\mathrm{wind}$ is the wind velocity,
and $r$ is the radius. The CSM density, which is the most influential parameter
in CSM interaction, is determined by $\dot{M}/v_\mathrm{CSM}$. In this paper, we
assume $v_\mathrm{wind}=100~\kmps$ and vary $\dot{M}$ to investigate the effects
of the interaction between SN~Ia ejecta and CSM. Our results are not 
significantly affected by the wind velocity as long as this is smaller than the
typical SN~Ia ejecta velocity which is much larger than $1000~\kmps$. Wind
velocities of SN~Ia progenitors are expected to be $\sim 10-1000~\kmps$
\citep[e.g.,][]{chomiuk2016}. The CSM extends to $10^{16}~\mathrm{cm}$ in our
standard models, so that the interaction continues through the peak of the
light-curve. The choice of CSM radius does not affect our conclusions as long as
the radius is sufficiently large, as discussed below. We also show some models
with different CSM radii. The CSM is assumed to have solar composition. The
exact composition, however, does not affect our results because the emission
properties are mostly determined by the hydrodynamics.
Fig.~\ref{fig:initial_density} shows the initial density structure of our
standard models. Our models cover from $\dot{M}=10^{-5}~\Msunpyr$ to
$10^{-3}~\Msunpyr$.

The major differences between our initial conditions and those in previous
studies of interaction between SN~Ia ejecta and CSM concern the density
structure. Previous studies considered a CSM formed in the mergers of two white
dwarfs and adopt a CSM density structure $\rho_\mathrm{CSM}\propto r^{-3}$
\citep[e.g.,][]{piro2016,maeda2018}. Thus, the radius of the dense CSM in those
studies is smaller than in our model. The CSM composition is also different, but
this is not expected to affect interaction-powered light curves, as they are
powered mechanically. 

\begin{figure*}
	\includegraphics[width=\columnwidth]{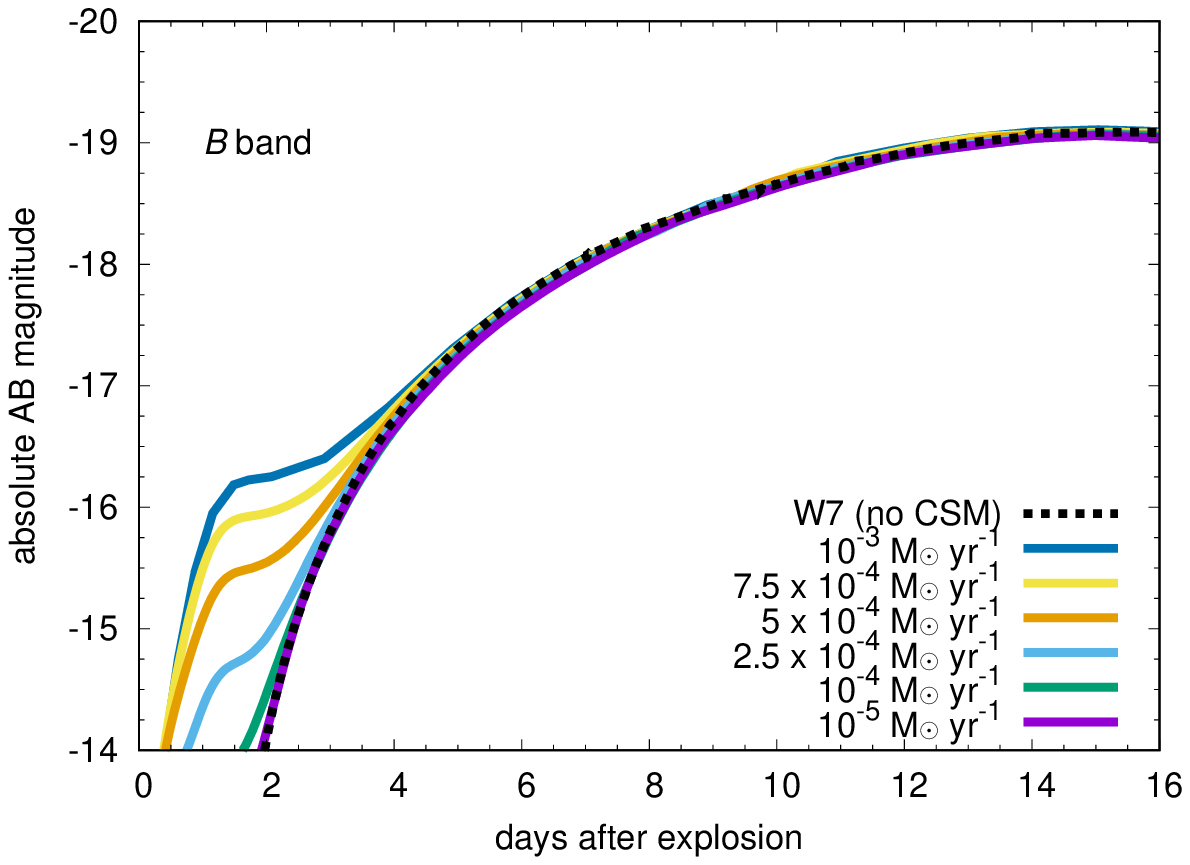}
	\includegraphics[width=\columnwidth]{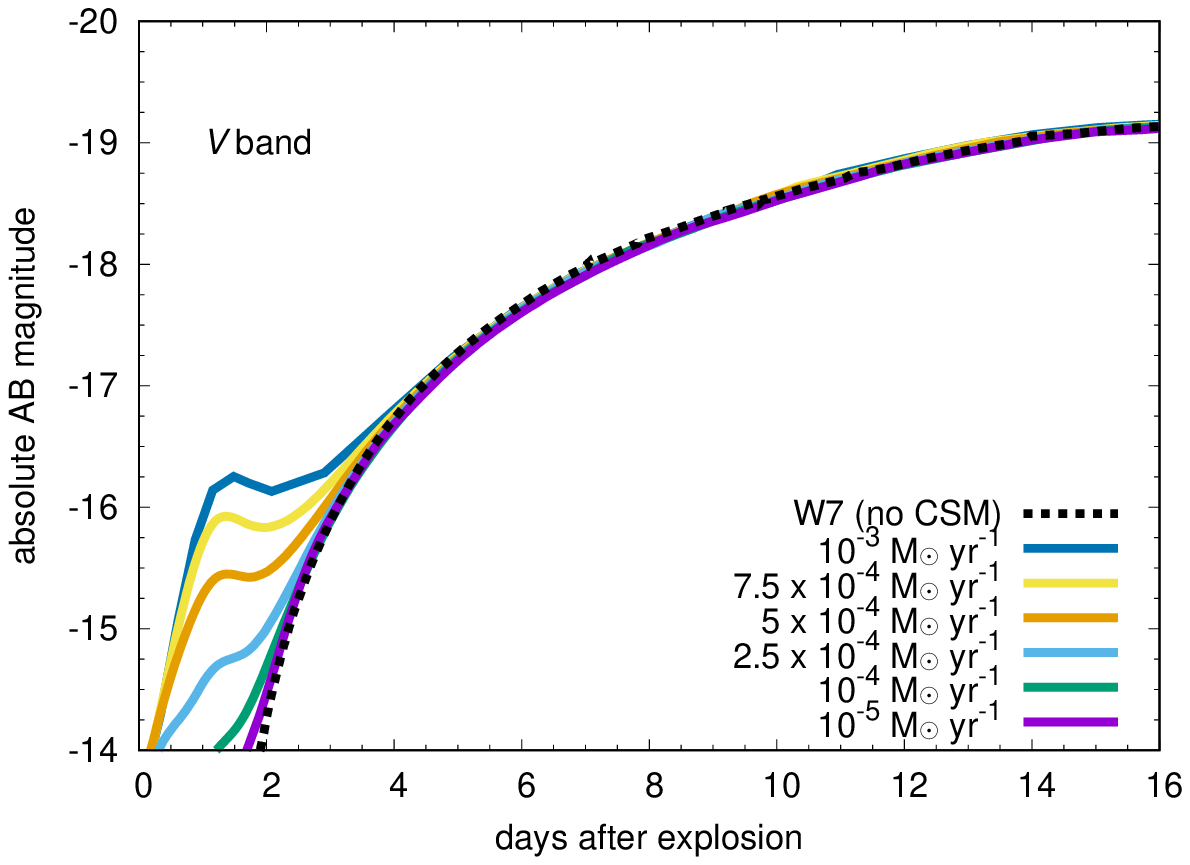}
	\includegraphics[width=\columnwidth]{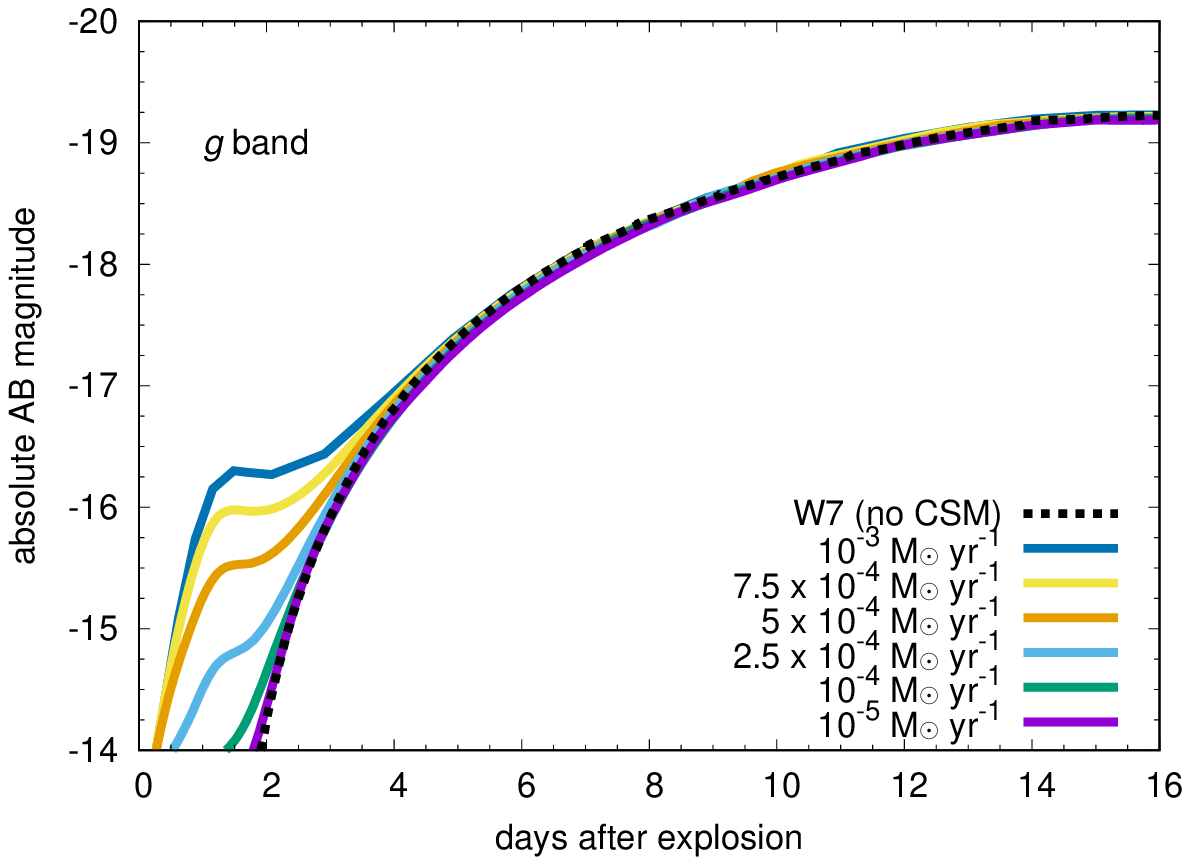}
	\includegraphics[width=\columnwidth]{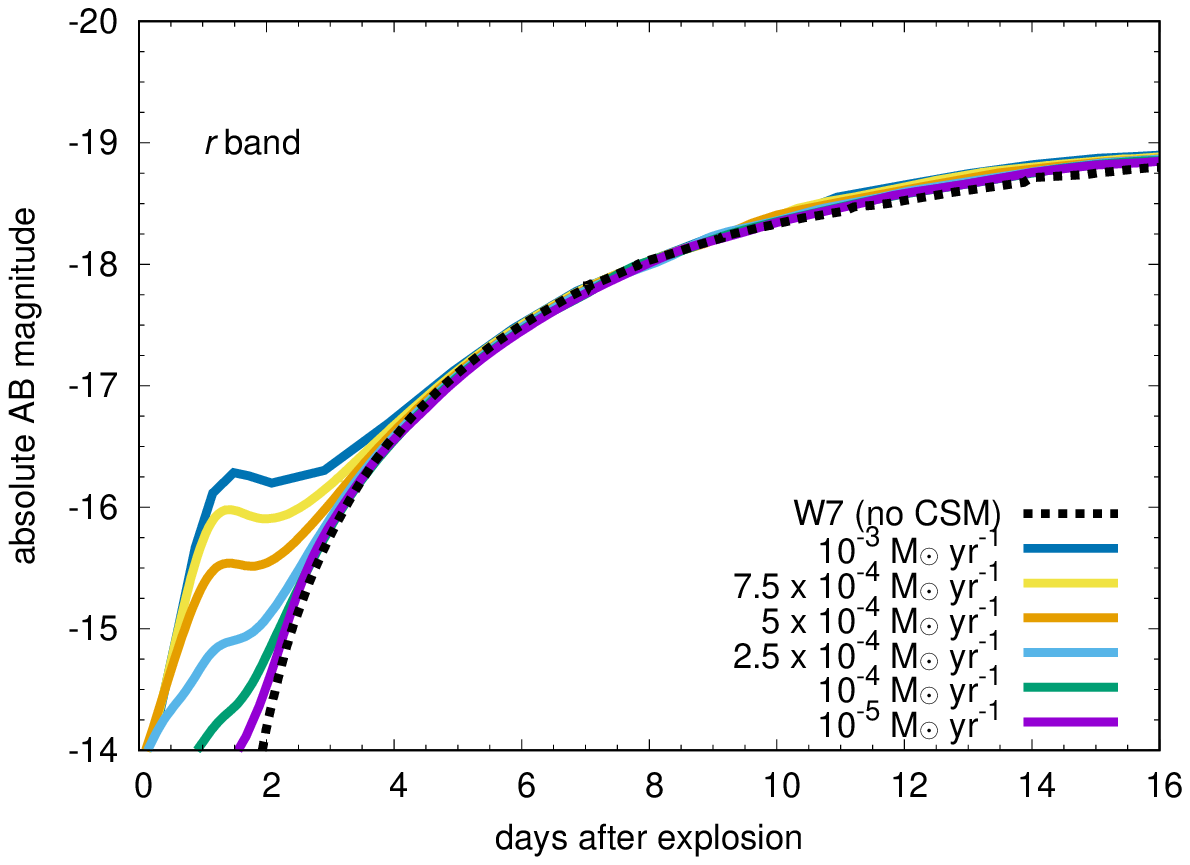}
    \caption{
    Optical light-curves obtained from our numerical calculations.
    }
    \label{fig:mag_opt}
\end{figure*}

\begin{figure}
	\includegraphics[width=\columnwidth]{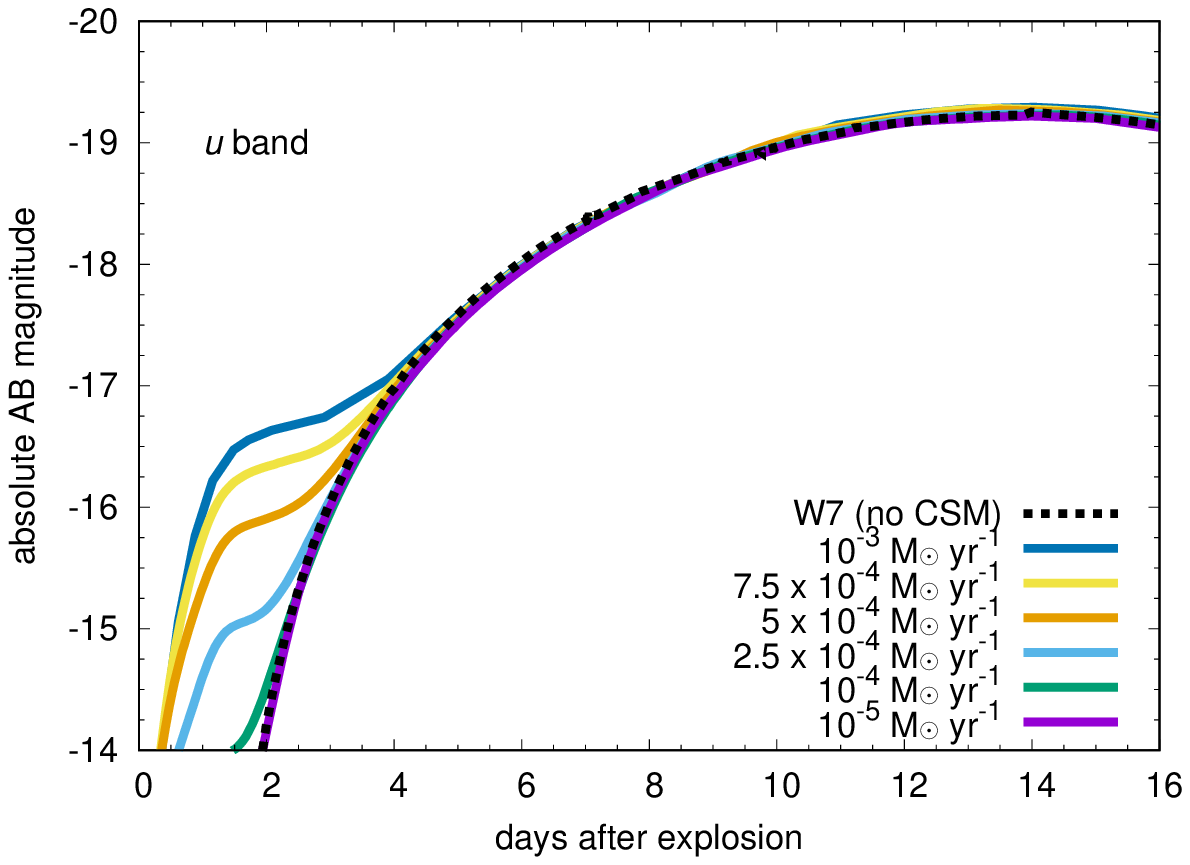}
	\includegraphics[width=\columnwidth]{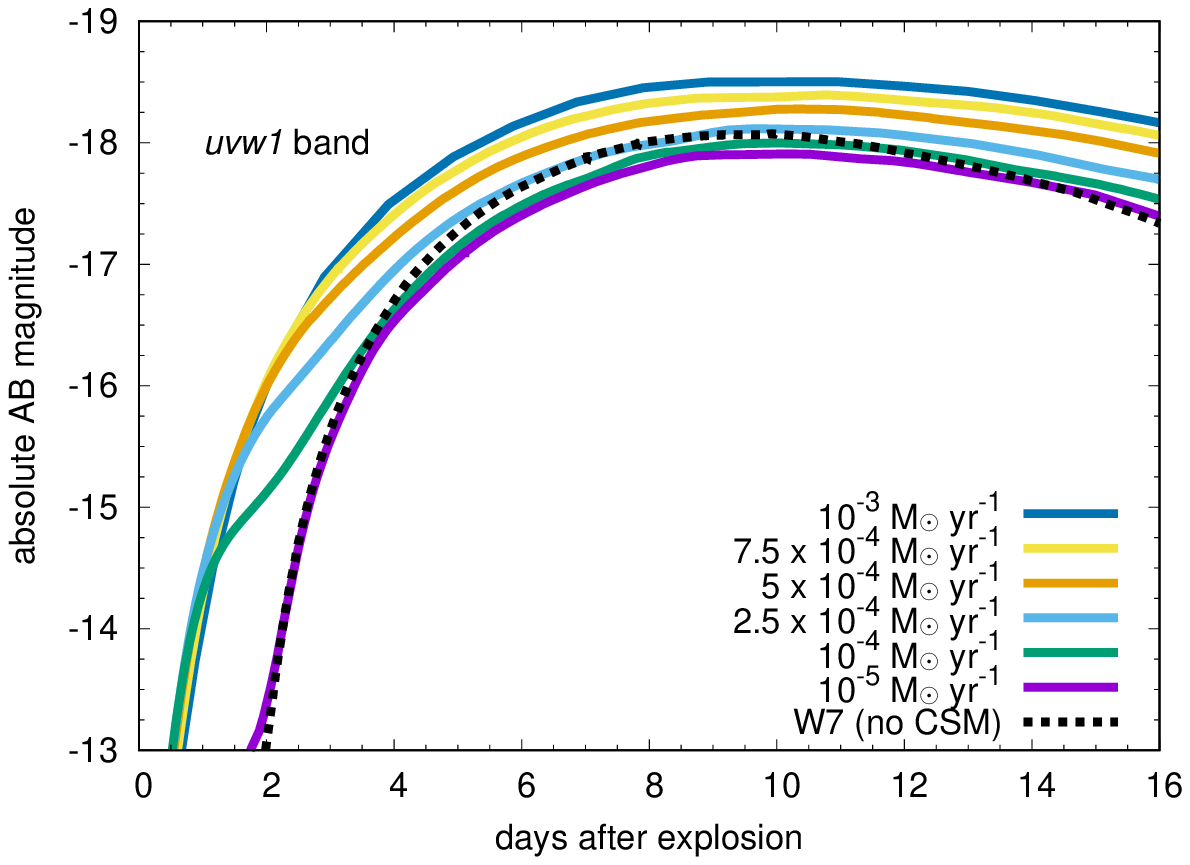}
	\includegraphics[width=\columnwidth]{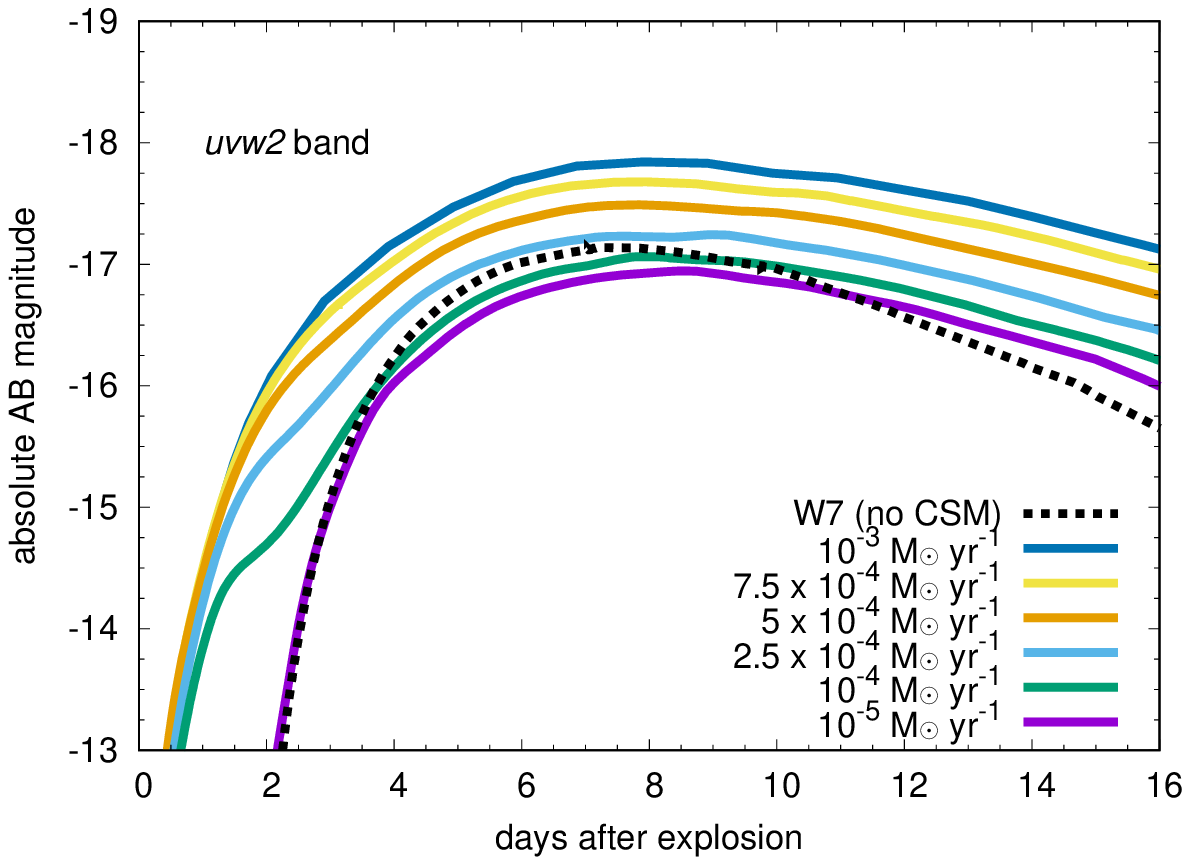}
    \caption{
    Ultraviolet light-curves obtained from our numerical calculations.
    }
    \label{fig:mag_uv}
\end{figure}

\section{Results}
\label{sec:results}

\subsection{Bolometric light-curves}

The top panel of Fig.~\ref{fig:bolometric} presents our synthetic bolometric
light-curves. We can find that the existence of CSM affects light-curve
properties of SNe~Ia. When $\dot{M}\lesssim 2.5\times 10^{-4}~\Msunpyr$, the
bolometric luminosity of SN~Ia models with CSM deviate from that without CSM
mostly at the earliest phases, within 4~days of explosion. Although CSM
interaction continues at later phases, the luminosity from CSM interaction is
much less than the luminosity caused by heating from \Ni\ and \Co\ decay. When
$\dot{M}\gtrsim 5\times 10^{-4}~\Msunpyr$, the luminosity provided by  CSM
interaction becomes sufficiently high and the bolometric luminosity is
higher than that in the model without CSM as long as the interaction continues.
Even in these cases, tthough, the contribution of CSM impact to the luminosity
is most significant at the earliest phases, within 4~days of explosion.

The bottom panel of Fig.~\ref{fig:bolometric} shows the pseudo-bolometric
luminosity evolution that is obtained by integrating the fluxes at optical
wavelengths ($3250-8900$~\AA). Compared to the bolometric light-curves, excess
luminosity due to CSM interaction is only noticeable within 4~days of explosion
in all the models. As discussed in the next section, the luminosity excess at
later epochs is mostly at ultraviolet (UV) wavelengths.  The effect of CSM
interaction is only visible at optical wavelengths when $\dot{M}\gtrsim
10^{-4}~\Msunpyr$, and only in the earliest phases. 

Fig.~\ref{fig:bolometric_radius} shows the effects of CSM radius on the
bolometric and pseudo-bolometric light-curves. The CSM is artificially cut at
$10^{16}~\mathrm{cm}$ in our models, but the actual outer edge of the CSM is
determined by the duration of the mass-loss phase from the progenitor system.
Although there are slight differences caused by the CSM radius, the effect of
CSM density is found to be much more significant than that of CSM radius. When
the CSM radius is too small, however, CSM interaction does not persist long
enough for a luminosity excess to be observed. To summarize, our conclusions in
this work are not affected by our assumption on the CSM radius if this is large
enough to cause CSM interaction at the earliest phases.

\subsection{Optical and ultraviolet light-curves}

Fig.~\ref{fig:mag_opt} shows our synthetic light-curves in optical bands. We
show the light-curves in the \textit{B} and \textit{V} bands from the Carnegie
Supernova Project (CSP, \citealt{hamuy2006}) and the \textit{g} and \textit{r}
bands from the Zwicky Transient Facility (ZTF, \citealt{bellm2019}). As
discussed in the previous section, optical light-curves are only affected at the
earliest phases within 4~days of explosion. We can find that CSM interaction
results in a significant early-phase flux excess in the optical bands when
$\dot{M}\gtrsim 10^{-4}~\Msunpyr$.

The light-curve evolution in UV bands below 3000~\AA\ is presented in
Fig.~\ref{fig:mag_uv}. We adopt the \textit{u} band ($3000-4000$~\AA) from CSP,
and the \textit{uvw2} and \textit{uvw1} bands from \textit{Neil Gehrels Swift
Observatory} \citep{gehrels2004,roming2005}, which are sensitive below 3000~\AA.
The \textit{u} band light-curves have similar properties to those of the optical
light-curves with the early-phase excess emission discussed above. However, UV
light-curves below 3000~\AA\ show significantly different light-curve features
from those found in optical light-curves. In particular, the UV light-curves
computed using $\dot{M}\gtrsim 5\times 10^{-4}~\Msunpyr$ show a persistent flux
excess at all phases when interaction is active. The SN UV flux below 3000~\AA\
is usually suppressed in SNe~Ia by absorption by Fe-group elements
\citep[e.g.,][]{hoflich1998,mazzali2000,lentz2000,kasen2006,mazzali2014,dessart2014b}.
Meanwhile, a large fraction of the radiation from CSM interaction is emitted at
the UV wavelengths because of the high shock temperature ($\sim 10^{4}$~K, e.g.,
\citealt{moriya2011}). This UV flux is emitted at large radii and can therefore
escape. Thus, the effect of CSM interaction is seen mainly at UV wavelengths,
and the extra flux from CSM interaction can easily dominate the UV flux of
interacting SNe~Ia.

\subsection{Colour}

\begin{figure}
	\includegraphics[width=\columnwidth]{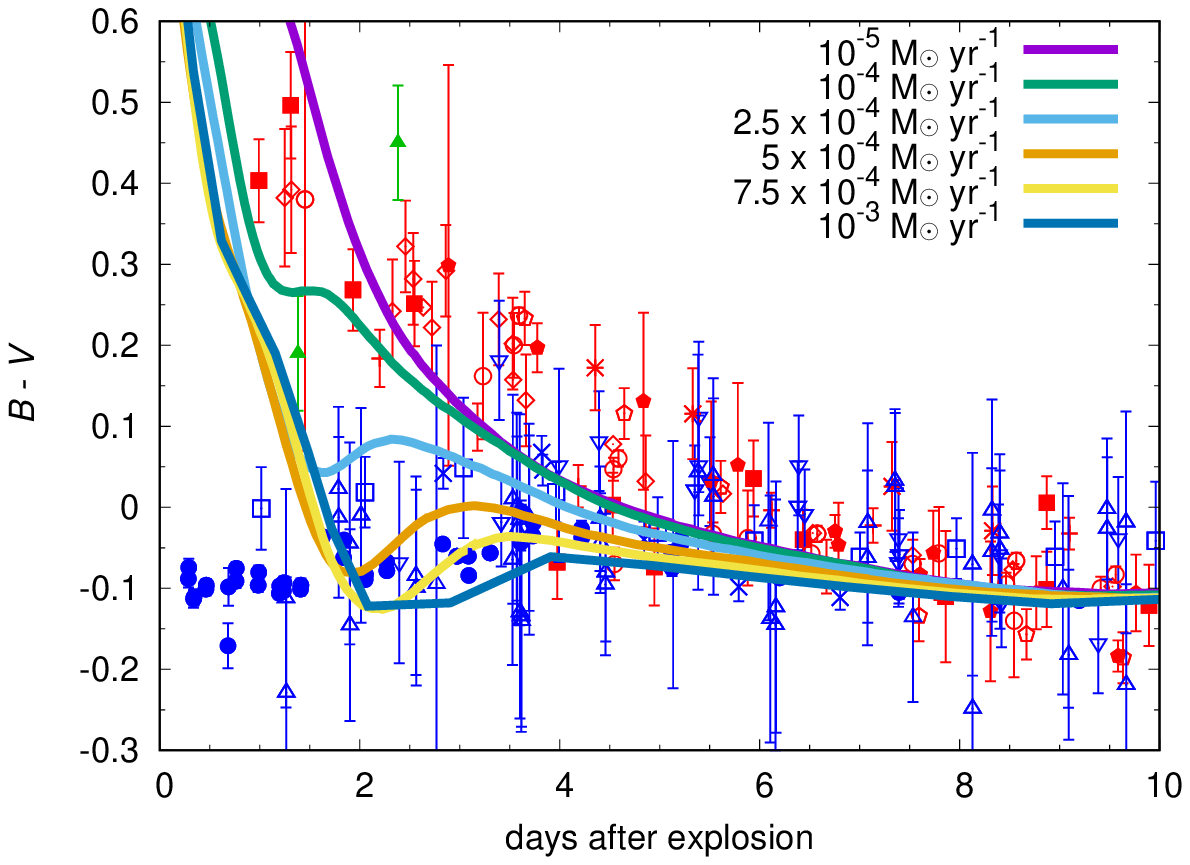}
	\includegraphics[width=\columnwidth]{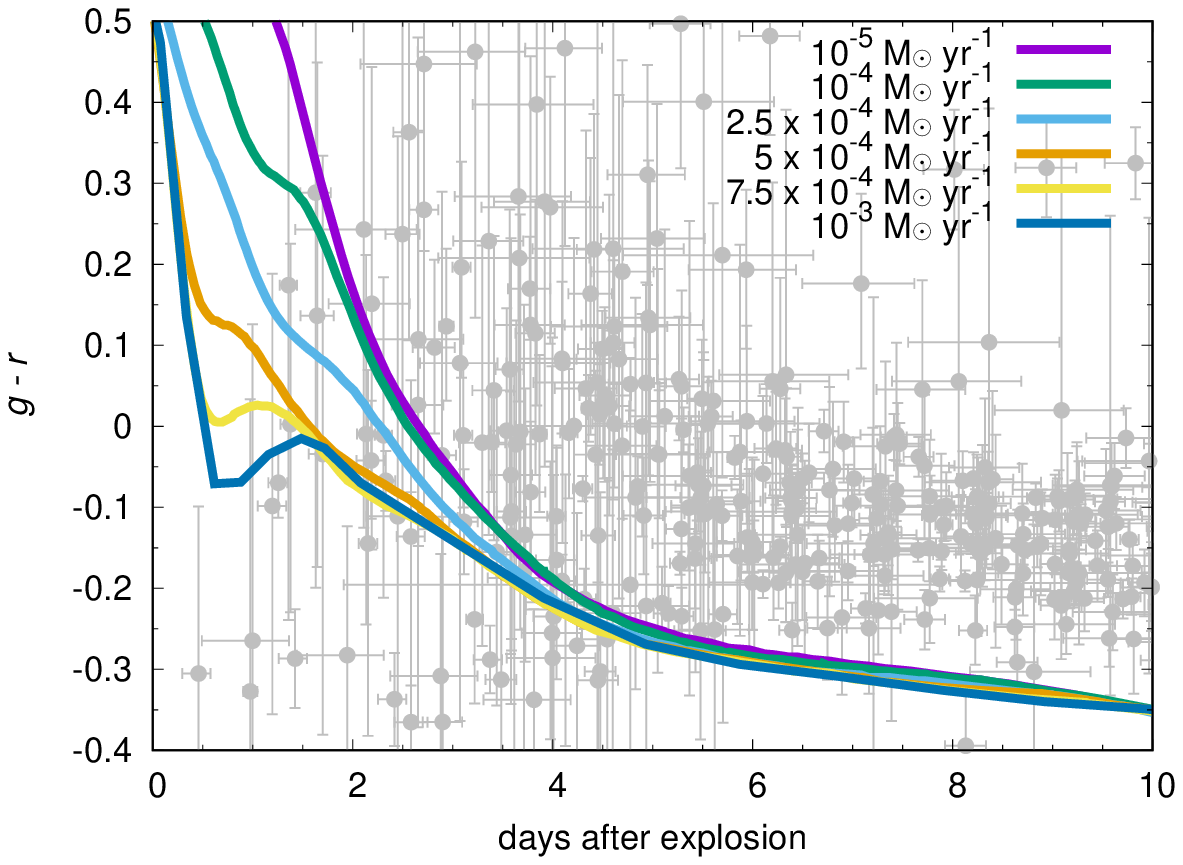}	
    \caption{
    Colour evolution of our synthetic light-curves. The top panel presents the $B-V$ colour evolution and the bottom panel presents the $g-r$ colour evolution. We also show the observed colour evolution of SNe~Ia from \citet[][top]{stritzinger2018} and \citet[][bottom]{bulla2020}. In the top panel, each SN~Ia is plotted with the same symbol as in \citet{stritzinger2018}, and the colour of the symbols is based on the early-phase behavior as defined in \citet{stritzinger2018}. The green symbol is the intermediate case. The details of each SN~Ia can be found in \citet{stritzinger2018}. SN~2017cbv is shown by blue circles.
    }
    \label{fig:colour}
\end{figure}

Fig.~\ref{fig:colour} shows that the colour evolution of our synthetic SN~Ia
models is also affected by CSM interaction. When the CSM density is high, SNe~Ia
are found to become blue more rapidly because of stronger CSM interaction. We
discuss a comparison with observations in the next section.

\begin{figure}
	\includegraphics[width=\columnwidth]{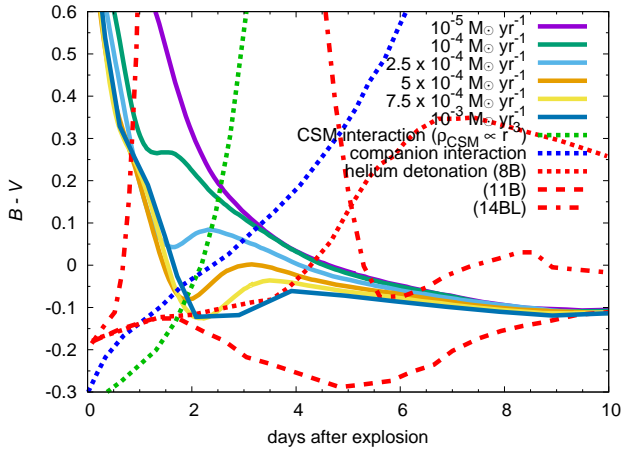}
    \caption{
Comparisons of the $B-V$ colour evolution among different theoretical models predicting early flux excess in SNe~Ia. Representative models in \citet{maeda2018} are presented.
}
    \label{fig:colour_modelcomparison}
\end{figure}

Figure~\ref{fig:colour_modelcomparison} compares the colour evolution of
several models predicting early flux excess in SNe~Ia from \citet{maeda2018}. A
CSM interaction model with a CSM density distribution of
$\rho_\mathrm{CSM}\propto r^{-3}$ from a WD merger, a companion
interaction model, and three helium detonation models (\Ni\ at outer layers) are
compared. Our CSM interaction model ($\rho_\mathrm{CSM}\propto r^{-2}$) is found
to have a distinct colour evolution compared to other models.

\begin{figure}
	\includegraphics[width=\columnwidth]{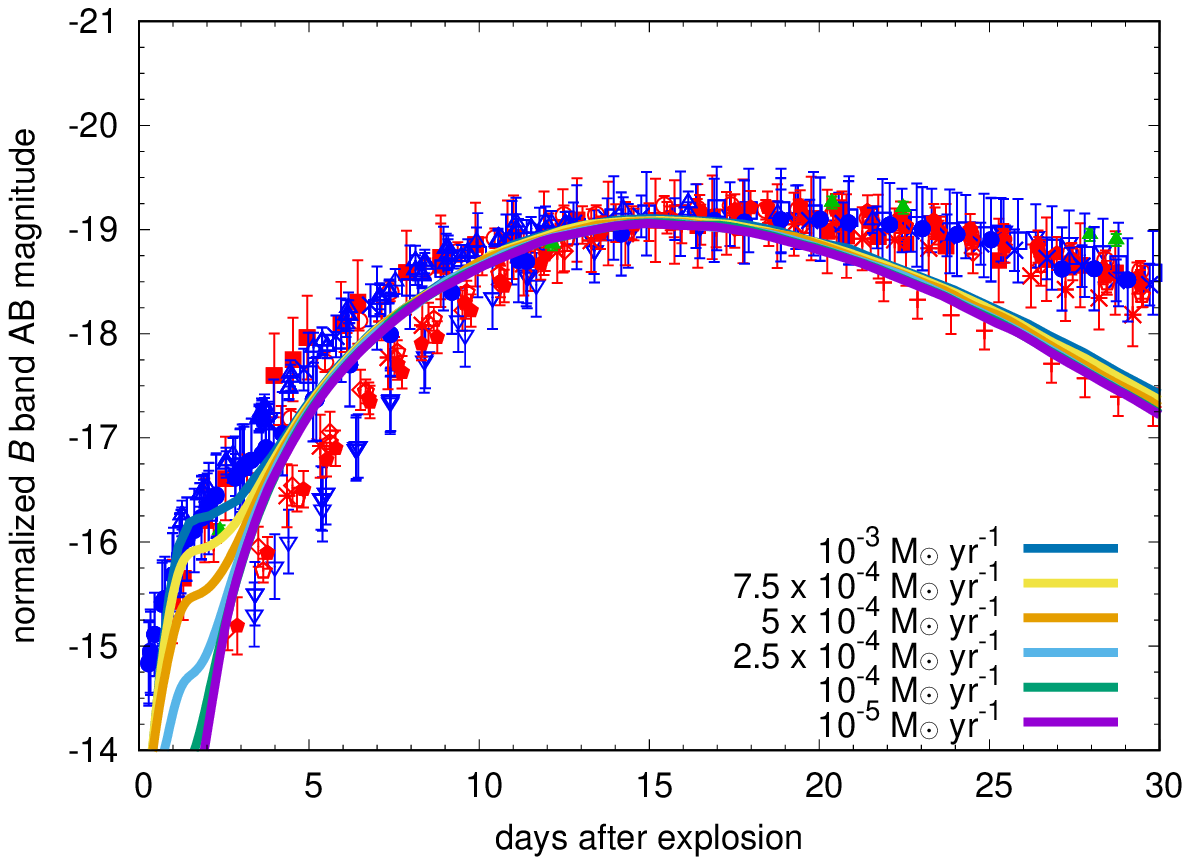}
	\includegraphics[width=\columnwidth]{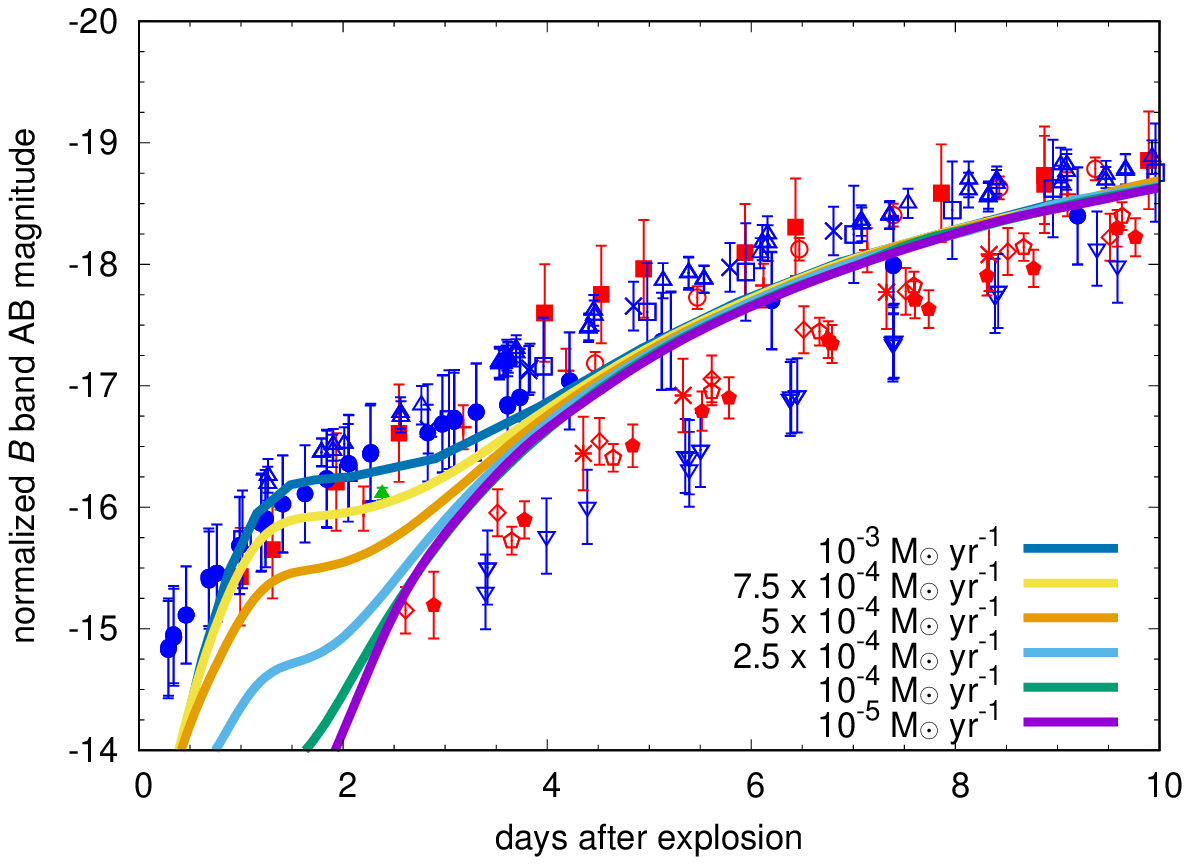}
    \caption{
    Scaled \textit{B} band light-curves of the SNe~Ia shown in the top panel of 
    Fig.~\ref{fig:colour}, which are taken from \citet{stritzinger2018}. The 
    observed magnitudes are scaled to match the synthetic models at  
    light-curve peak. The top panel shows the light-curves until 30~days after 
    explosion and the bottom panel shows the light-curves until 10~days after 
    explosion. SN~2017cbv is shown as blue circles.
    }
    \label{fig:lightcurveB}
\end{figure}

\section{Discussion}
\label{sec:discussion}

We have shown that the collision between SNe\,Ia ejecta and an extended CSM 
caused by a steady wind from the progenitor system leads to a flux excess in the
SN optical light-curves in the first 4~days after explosion if $\dot{M}\gtrsim
10^{-4}~\Msunpyr$ with a wind velocity of 100~\kmps\ (Fig.~\ref{fig:mag_opt}).
The colour evolution is also affected by CSM interaction
(Fig.~\ref{fig:colour}). It has been suggested that there may be distinct
populations in colour evolution in the early phases ($\gtrsim 4~\mathrm{days}$)
of SNe~Ia \citep{stritzinger2018}, although some studies find the evidence less
conclusive \citep[e.g.,][]{bulla2020}. In order to test whether interaction with
a CSM could lead to the observed behaviour, the observed colour evolution of
SNe\,Ia is compared to the colour evolution in our synthetic models in
Fig.~\ref{fig:colour}. In the case of the $B-V$ colour evolution presented by
\citet{stritzinger2018}, we find that SNe~Ia with an early red colour 
are consistent with the models with $\dot{M}\lesssim
10^{-4}~\Msunpyr$, while SNe~Ia with an early blue colour evolution are more
consistent with models with $\dot{M}\gtrsim 2.5\times10^{-4}~\Msunpyr$. The
change from red evolution to blue evolution occurs over a small range of
$\dot{M}$ ($10^{-4}~\Msunpyr\lesssim \dot{M}\lesssim 2.5\times
10^{-4}~\Msunpyr$).

The ZTF SNe\,Ia sample \citep{bulla2020} does not show the presence of two
groups with different $g-r$ colour evolution (Fig.~\ref{fig:colour}). Our
synthetic models match the $B-V$ evolution of the SNe~Ia sample in
\citet{stritzinger2018} well, from early to late phases, but the $g-r$ colour
evolution is quite different at most phases from that of the ZTF SNe~Ia as
reported by \citet{bulla2020}. We cannot offer a clear reason for this
difference. As our numerical calculations do not resolve spectral line features
in the \textit{g} and \textit{r} bands well, some lines in spectra might be
responsible for the difference.
It is also possible that the explosion dates of the ZTF SN~Ia sample, which
are estimated by adopting a power-law function, are not well determined, and
there might be a systematic shift in the colour evolution.

If the separation in $B-V$ colour evolution at early phases is the result of  a
difference in CSM, SNe~Ia that are blue at the early phases should always have
an early flux excess from CSM interaction. Fig.~\ref{fig:lightcurveB} shows the
$B$ band light-curves of SNe~Ia in the Stritzinger's sample (top panel of
Fig.~\ref{fig:colour}). SNe~Ia with an early \textit{B}-band flux excess tend to
have blue colour. In particular, the flux excess
and early colour evolution of SN~2017cbv
\citep{hosseinzadeh2017,stritzinger2018} are well explained by our CSM
interaction model. However, one SN~Ia (SN~2009ig,
\citealt{blondin2012,foley2012,marion2013}) with early flux excess has red
colour evolution. This suggests that more than one mechanism may be responsible
for the early flux excess, as discussed in Section~\ref{sec:introduction}.
The colour evolution of
SN~2017cbv towards the blue is faster than the model prediction.
It may be an effect of asphericity in the CSM, allowing photons from
the interaction region to escape faster. The discrepancy between the
synthetic and observed light curves after peak seen in
Fig.~\ref{fig:lightcurveB} may be caused by differences inside the ejecta
and it does not
affect the early bumps caused by the interaction with the outermost ejecta.

An early flux excess due to CSM interaction requires mass-loss rates from SN~Ia
progenitor systems of at least $\dot{M}\gtrsim 10^{-4}~\Msunpyr$, assuming a
wind velocity of 100~\kmps. Such high mass-loss rates are usually excluded by
radio and X-ray observations (Section~\ref{sec:introduction}). However, we do
know that some SN\,Ia-CSM progenitor systems had mass-loss rates exceeding
$10^{-3}~\Msunpyr$ \citep[e.g.,][]{sharma2023}. It is possible that SN Ia
progenitor systems with $10^{-4}~\Msunpyr\lesssim \dot{M}\lesssim
10^{-3}~\Msunpyr$ do not exhibit clear CSM interaction signatures in spectra as
do SNe~Ia-CSM, but only show a flux excess in the light-curves.

SNe~Ia with a moderate mass-loss rate such that they only show the early flux
excess in optical bands but no spectroscopic signature may still show
significant, persistent flux excess also in the UV, and CSM interaction may thus
explain some diversity in the UV properties of SNe~Ia
\citep[e.g.,][]{brown2012,brown2020,pan2018,pan2020,srivastav2022,sauer2008}.
Because $20-30$\% of SNe~Ia show early flux excess
\citep[e.g.,][]{deckers2022,magee2022} and there are several mechanisms to cause
early flux excess, up to around 10\% of SNe~Ia may have such early flux excess
due to CSM interaction discussed in this paper. \citet{chomiuk2016}
estimated that less than 10\% of SNe~Ia have mass-loss rates higher than
$10^{-4}~\Msunpyr$ based on radio luminosity limits. This is consistent with our
estimates of the fraction of SNe~Ia with early flux excess from wind
interaction. SN~Ia progenitor systems with high mass-loss rates may
result from optically thick winds \citep[e.g.,][]{kato1994} or common-envelope winds \citep[e.g.,][]{cui2022}, for example.
The fraction of SNe~Ia with early flux excess from
the CSM interaction is presumably 
higher than
the fraction of SNe~Ia-CSM among SNe~Ia ($0.02-0.2$\% of SNe~Ia,
\citealt{sharma2023}).

Finally, we note that SNe~Ia without early flux excess show a rather slow
light-curve rise, often referred to as a ``dark phase''
\citep[e.g.,][]{mazzali2001,piro2013}.
Fitting bolometric light curves with a
$t^2$ function leads to an explosion date estimate of about one day later than
the actual explosion date (Fig.~\ref{fig:bolometric}). A difference of even just
one day may be critical in interpreting the earliest photometric and
spectroscopic properties of SNe~Ia. Interpretation of early observational data
needs a careful assessment of the explosion date.

\section{Conclusions}
\label{sec:conclusions}

We investigated the effect on the photometric properties of SNe~Ia of the
interaction between the SN ejecta and a surrounding hydrogen-rich, dense,
extended CSM. The CSM is assumed to be formed by steady mass loss from the
progenitor system. It is assumed to have an extended structure with
$\rho_\mathrm{CSM}\propto r^{-2}$ (Eq.~\ref{eq:csmdensity}) and solar
metallicity. An early flux excess in optical light-curves is caused by CSM
interaction if $\dot{M}\gtrsim 10^{-4}~\Msunpyr$.  UV (below 3000~\AA)
light-curves are affected significantly and the flux excess there persists as
long as CSM interaction continues.

CSM interaction also affects the early colour evolution in SNe~Ia. When the CSM
density is high enough to cause an early optical flux excess, the optical colour
is also found to evolve rapidly towards the blue. Thus, CSM interaction may be
partly responsible for the diversity of the early colour evolution that has been
seen in SNe~Ia. It is possible that a fraction of SNe~Ia do not have high enough
CSM density for the spectra to display the features of  SNe~Ia-CSM
($\dot{M}\gtrsim 10^{-3}~\Msunpyr$), but still have a sufficient CSM density to
cause early-phase flux excess in optical light-curves. This requires  
($10^{-4}~\Msunpyr\lesssim \dot{M}\lesssim 10^{-3}~\Msunpyr$). CSM interaction
should also affect the X-ray and radio properties. Multi-wavelength observations
of SNe~Ia is therefore important in constraining their CSM properties.  Such
high mass-loss rates are rather extreme even among single-degenerate progenitor
systems. However, even though systems with such high mass-loss rates may be
rare, they can provide important clues as to the evolutionary path towards
SNe~Ia. Because the CSM interaction signatures appear clearly in the UV, early
SN~Ia observations by future ultraviolet satellites such as
ULTRASAT\footnote{\url{https://www.weizmann.ac.il/ultrasat/}} and
CASTOR\footnote{\url{https://www.castormission.org/}} will be critical in
uncovering the circumstellar environments and progenitor systems of SNe~Ia.

\section*{Acknowledgements}
We thank Benjamin Shappee and Maximilian Stritzinger for useful discussions.
This work was supported by the NAOJ Research Coordination Committee, NINS (NAOJ-RCC-22FS-0502, NAOJ-RCC-22FS-0503).
PAM and EP are grateful for support and hospitality at NAOJ in Mitaka under program NAOJ-RCC-22FS-0502 and NAOJ-RCC-22FS-0503.
TJM is supported by the Grants-in-Aid for Scientific Research of the Japan Society for the Promotion of Science (JP20H00174, JP21K13966, JP21H04997).
AC acknowledges support by NASA grant JWST-GO-02114.032-A.
Numerical computations were in part carried out on PC cluster at Center for Computational Astrophysics (CfCA), National Astronomical Observatory of Japan.

\section*{Data Availability}
The data underlying this article will be shared on reasonable request to the corresponding author.



\bibliographystyle{mnras}
\bibliography{mnras} 







\bsp	
\label{lastpage}
\end{document}